\documentclass{aa}
\usepackage{graphicx}

\newcommand{\lsun}{L$_\odot$}
\newcommand{\msun}{M$_\odot$}
\begin{document}
\title{
The supernova rate per unit mass
}


\author{
     F. Mannucci,    \inst{1}
     M. Della Valle, \inst{2,3}
     N. Panagia,     \inst{3,4}
     E. Cappellaro,  \inst{5}
     G. Cresci,      \inst{6}
     R. Maiolino,    \inst{2}
     A. Petrosian,   \inst{7}
\and M. Turatto      \inst{8}
}

\offprints{F. Mannucci \email{filippo@arcetri.astro.it}}

\institute{
     CNR - IRA, Largo E. Fermi 5, 50125 Firenze, Italy
\and INAF, Osservatorio Astrofisico di Arcetri, Largo E. Fermi 5, 50125
     Firenze, Italy
\and Space Telescope Science Institute, 3700 San Martin Drive, 
     Baltimore, MD 21218, USA 
\and on assignment from the Space Telescope Operations Division, Research and
     Scientific Support Department of ESA
\and INAF, Osservatorio Astronomico di Capodimonte, salita Moiariello a
     Capodimonte 16, 80131 Napoli, Italy
\and Dipartimento di Astronomia, Universit\'a di Firenze, 
	 Largo E. Fermi 5, I-50125, Firenze, Italy  
\and Byurakan Astrophysical Observatory and Isaac Newton Institute of 
     Chile, Armenian Branch, Byurakan 378433
\and INAF, Osservatorio Astronomico di Padova, vicolo dell'Osservatorio 5,
     35122 Padova, Italy
}

\authorrunning{F. Mannucci et al.}
\titlerunning{The supernova rate per unit mass}

\date{}

\abstract{We compute the rate of supernovae (SNe) of different types
along the Hubble sequence normalized to the near-infrared luminosity
and to the stellar mass of the parent galaxies. This is made possible
by the new complete catalog of near-infrared galaxy magnitudes
obtained by 2MASS.  We find that the rates of all SN types, including
Ia, Ib/c and II, show a sharp dependence on both the morphology and
the (B--K) colors of the parent galaxies and, therefore, on the
star formation activity.  In particular we find, with a high
statistical significance, that the type Ia rate in late type galaxies
is a factor $\sim$20 higher than in E/S0. Similarly, the type Ia rate
in the galaxies bluer than B--K=2.6 is about a factor of 30 larger
than in galaxies with B--K$>$4.1.  These findings can be explained by
assuming that a significant fraction of Ia events in late
Spirals/Irregulars originates in a relatively young stellar component.

}

\maketitle

\section{Introduction}

The supernova (SN) rate normalized to the stellar mass of the
parent galaxies contains unique information on the initial
mass function of stars in the range of masses between about 3 and
100 $M_\odot$ (e.g. Madau, Della Valle \& Panagia, 1998), 
therefore it is a very
powerful tool for understanding the formation and the chemical
evolution of the galaxies and constraining their star formation
histories.  In particular, the rate of the so called ``core-collapse''
(CC) supernovae, i.e. type II and Ib/c, which have massive progenitors
(e.g. Woosley, Heger \& Weaver, 2002), reflects the instantaneous birth
rate of stars more massive than 8$M_\odot$ (e.g. Iben \& Renzini
1983), whereas the trend of the frequency of type Ia SNe
from Ellipticals to late Spirals can provide deep insights on the
controversial nature 
of the binary companion in type Ia events
(see for example Hamuy et al. 2003 and Livio \& Riess 2003).
In this
paper we focus our attention on the dependence of the SN rate
on the morphological Hubble type and on the B--K color of the parent
galaxies. The empirical grounds for this kind of study are provided by
the supernova surveys carried out in the local Universe in the past
years (e.g. Cappellaro et al. 1999, hereafter C99) or still ongoing,
such as LOTOSS (Filippenko et al. 2001). Systematic
surveys allow to compute the SN rates by applying the control time
technique (Zwicky 1942) after taking into account the various
selection effects (e.g. Cappellaro et al. 1997, hereafter C97). 
In all cases, the SN
rate is normalized to some quantity somehow related
to the galaxy ``sizes''.  The quantity most commonly used is the luminosity
in the optical B band (e.g. van den Bergh \& Tammann 1991, Tammann et
al., 1994), assumed to be a measure of the stellar mass at least for galaxies 
of the same morphological class (Tammann, 1974).  Thus, the classical
SN unit (SNuB) is defined as number of events per century per
$10^{10}$\lsun\ in the B band.  
More recently,  C99 and Mannucci et al. (2003)  used the 
far-infrared
luminosity, commonly considered to be proportional to the Star
Formation Rate (SFR)  (e.g., Hirashita, Buat \& Inoue, 2003), to
normalize the SN rate thus expressing the rates in SNuIR.

In their most recent determination of the local rates C99 found:
{\sl a)} the rates of types Ib/c and II SNe are null in the E/S0
galaxies and show a moderate increase from Sa to Sd types; {\sl b)} a
marginally significant decrease of type II rates is also observed from
Sd to Irr.  {\sl c)} the rate of type Ia SNe measured in SNuB is
almost constant along the Hubble sequence from Elliptical to Sd
galaxies and shows a modest increase toward the Irregular and peculiar
galaxies (admittedly with a low statistical significance).  The
interpretation of the latter result is not obvious: if type Ia SNe are
related to old stellar populations, one would expect a decreasing rate
through the whole morphological Hubble sequence from E/S0 to
Irr. The problem may be due to the normalization to the B 
luminosity which
is a poor tracer of the stellar mass along the whole Hubble sequence
(see section~\ref{sec:snmass}).

In this paper we intend to go
beyond this approach and derive the rates of the various types of
SNe normalized to the stellar mass of the parent galaxies as 
inferred from the K-band luminosity.
This is now possible thanks to the Two Micron All Sky Survey (2MASS
\footnote{The Two Micron All Sky Survey is a joint project of the 
University of Massachusetts and the Infrared Processing and Analysis 
Center/California Institute of Technology, funded by the National 
Aeronautics and Space Administration and the National Science 
Foundation.}, Jarrett et al., 2003) whose catalog of
near-infrared magnitudes of extended objects is now complete. In
section~\ref{sec:sncatalog} the input catalog will be presented, in
sections~\ref{sec:kmag} and \ref{sec:mass} the near-infrared
magnitudes and the recipe to compute the mass will be discussed.  In
the next sections we will present the results and their
interpretation.

\section{Importance of the SN rate per unit mass}
\label{sec:snmass}

In the past decades the luminosity in the B band has been used as
a rough gauge of the mass of the galaxies. Indeed B luminosity has been for
long time the only available photometric measurement for most systems
observed in the local Universe. Since the presence of a young stellar
component can contribute significantly to the B luminosity in the late
Hubble type galaxies, the proportionality between B luminosity and
stellar mass is expected to change dramatically along the Hubble
sequence. As an example, we note that, after using the galaxy colors
by Fioc \& Rocca-Volmerange (1999) to compute the M/L ratio as in Bell
\& de Jong (2001, hereafter BJ01), a difference of a factor of 10 is
found in the stellar mass between an elliptical and an irregular
galaxy with the same B magnitude and colors typical of their
classes. Large differences can also be found within each class of
galaxies: as an example, a range of a factor of 3 in mass is found for
Sb galaxies having (B--K) colors within $\pm1\sigma$ the average of
their class.

In addition we note that 
the B flux is the result of combined emission from
old stars, emission from young populations and absorption by dust,
with the relative contributions changing along the Hubble sequence. Even
if B light would be an acceptable measure of the stellar mass in the E/S0
galaxies, it is a very poor tracer of mass along the whole
Hubble sequence.

In the past years, a number of authors (van den Bergh 1990; Della
Valle \& Livio 1994; Panagia 2000) 
had normalized the rate of type Ia SNe to the near-infrared H and K bands, 
which are
better tracers of stellar mass than B light. They found a sharp
increase of the production of type Ia SNe toward late morphological
types  (see also Mannucci et al. 2003). However, in those the
computation of the normalized SN rate was carried out by using 
average colors for the parent galaxies, because at that time the
individual near-infrared measurements were not available for most galaxies.
Recently, the 2MASS collaboration released their catalog 
(Jarrett et al., 2003), and now an accurate normalization 
of the SN rate to the total stellar mass has eventually become possible.

\section{SN catalog}
\label{sec:sncatalog}

In this study we used the SN catalog from C99. This sample is a
compilation of 136 SNe \footnote{One of the 137 SNe used in the C99
was later discovered to be associated to a galaxy not 
included in the sample} discovered by
five groups: the SN search of the Padova group with the Asiago Schmidt
telescopes, the Sternberg Institute search at the Crimea observatory,
the visual search by R. Evans, the survey by the Observatoire of the
C\^ote d'Azur, and the Calan/Tololo search.  The
full descriptions of these searches and the relative references are
given in C97.  These searches
were chosen because all the information for the computation of the
control time and of the observational bias, such as the 
galaxy sample, the
frequency of the observations and the limit magnitudes, were
available.  Only SNe in galaxies of known luminosity, colors,
morphology, inclination and distance can be considered in our study.
Therefore, even if the merging of the five original catalogs contains
about 250 SNe, the sample used by C99 includes only 136 events
occurred in galaxies reported in the RC3 catalog (de Vaucouleurs et
al, 1991) and in the LEDA database \footnote{the Lyon-Meudon
Extragalactic Database (LEDA) is supplied by the LEDA team at the
CRAL-Observatoire de Lyon (France)}.
The SNe in the final catalog have an average distance of 38 Mpc, 
with 90\% of them below 85 Mpc.

The computational procedure to correct for the selection effects
is described in C97 and is based on several
ingredients as the absolute magnitude at maximum, the
light evolution in the B band, 
the average extinction, the correction
for inclination of the spiral galaxy with respect to the line of sight,
the fraction of SNe lost in the overwhelming brightness of the
galaxy nuclei. For consistency, we used the same assumptions and
procedures as in C97 and C99. 

\section{The near-infrared magnitudes}
\label{sec:kmag}

\begin{figure}	
\centering
\includegraphics[width=9cm]{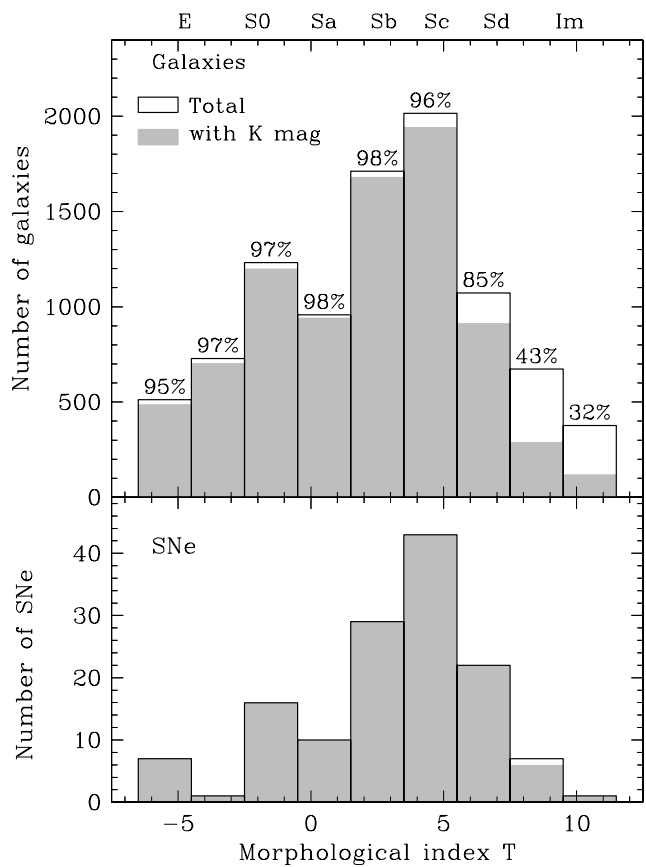}
\caption{
{\em Upper panel:} The histogram of the number of galaxies
in the input C99 catalog (in white) as a function of the
morphological index T
is compared with that of the galaxies having a
2MASS detection in K (in gray).
The fraction of detected galaxies is shown over each bin.
{\em Lower panel:} histogram of the number of SNe
as a function of the morphological index of the parent galaxy.
In white the input C99 catalog, in grey the catalog with retrieved K-band
magnitude.
}
\label{fig:missing}
\end{figure}

\begin{figure}	
\centering
\includegraphics[width=9cm]{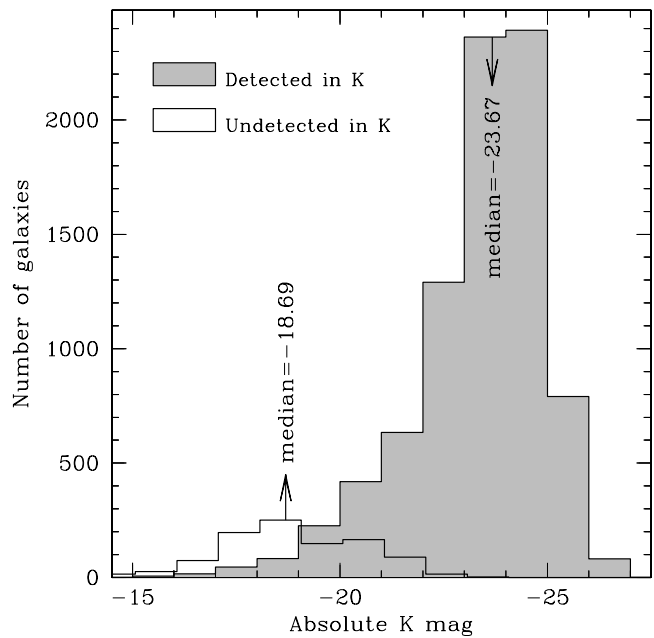}
\caption{The absolute K-band magnitude of the galaxies detected by 2MASS
is compared to that of the undetected galaxies (see text). 
It is apparent that most
the of galaxies missing from the 2MASS catalog are local dwarfs contributing
very little to the total K-band luminosity and mass. The
difference between the median of the two distribution is about 5 magnitudes.}
\label{fig:misslum}
\end{figure}

We cross-correlated the catalog of the 9346 galaxies of C99 
with the 2MASS Extended Source Catalog (XSC) at IRSA 
\footnote{http://irsa.ipac.caltech.edu} to obtain the near-IR magnitudes.  
The coordinates were
obtained from the HyperLeda catalog
\footnote{http://leda.univ-lyon1.fr}
(Paturel et al., 1989; Prugniel \& H\'eraudeau, 1998). 
Among the possible choices of photometric measurements in
the 2MASS catalog, we used the total photometry obtained by
extrapolating the fit of a Sersic function 
to the galaxy radial profile 
({\em k\_m\_ext} in the 2MASS nomenclature, see Jarrett et al., 2003, for
details).  This is the photometric
measurement best suited to represent the total magnitude to be compared
with the RC3 total B magnitudes.  We retrieved 8349 galaxies present
in the 2MASS XSC, i.e., 89\% of the input catalog.
Figure~\ref{fig:missing} shows the distribution of 
the morphological indexes of the galaxies with available K
magnitudes compared to the
distribution of the original C99 catalog (upper panel).  
It shows that the 2MASS
catalog contains the near-infrared magnitudes of essentially all 
the galaxies with morphological types
between E and Sd, while about 2/3 of the Irr are missing. Most of them
are actually present at a faint level in the 2MASS images, at least in J, 
but are not retrieved by the XSC software.

We can estimate an upper limit to the contribution to the total K-band
luminosity due to the galaxies not detected by 2MASS assuming they all
have a K-band magnitude equal to the survey limit for extended objects
of K=13.5 (see www.ipac.caltech.edu/2mass/overview/about2mass.html).
Figure~\ref{fig:misslum} shows the
resulting absolute magnitudes: most of the galaxies undetected by
2MASS are local dwarf galaxies, small systems which are bright in the
B band but faint in the K band.  Their contribution to the total
K-band luminosity of the total galaxy sample is below the tiny
fraction of 0.3\%, and it is below 5\% even if only the Irr galaxies
are considered. For this reason the incompleteness of the 2MASS XSC
catalog is not a serious problem for the computation of the rates.\\

This is confirmed by the fact that only 1 SN out of the input list 
of 136 events occurred in a galaxy
not present in the 2MASS catalog. Such a SN was, therefore, 
removed from the sample (see Figure~\ref{fig:missing}, lower panel). 

For 64 galaxies (0.77\% of the sample) the morphological
classification is absent or very uncertain. These galaxies
were excluded from the sample when computing the SN rate as a function of the
morphology (section~\ref{sec:snrate}) but were
used to compute the SN rate as a
function of the B--K color (section~\ref{sec:sncolor}).

The final numbers of SNe and galaxies 
used in our computations are reported in Table~\ref{tab:snn}.
The non integer number of SNe are due to the presence of a few
SNe with incomplete or unknown classification which were 
divided into different classes, as explained in C97.

\begin{table}
\caption{Number of SNe and galaxies per morphological bin}
\begin{tabular}{lrrrr}
\hline
\hline
Type  & Ngal &  Ia~  &  Ib/c  & II~  \\
\hline
E/S0  & 2048 &  21.0 &    0  &   0  \\
S0a/b & 2911 &  18.5 &  5.5  & 16.0 \\
Sbc/d & 2682 &  21.4 &  7.1  & 31.5 \\
Irr   &  644 &   6.8 &  2.2  & 5.0  \\
\hline
\end{tabular}
\label{tab:snn}
\end{table}

\section{Galaxy mass}
\label{sec:mass}

  \begin{figure}	
  \centering
  \includegraphics[width=9cm]{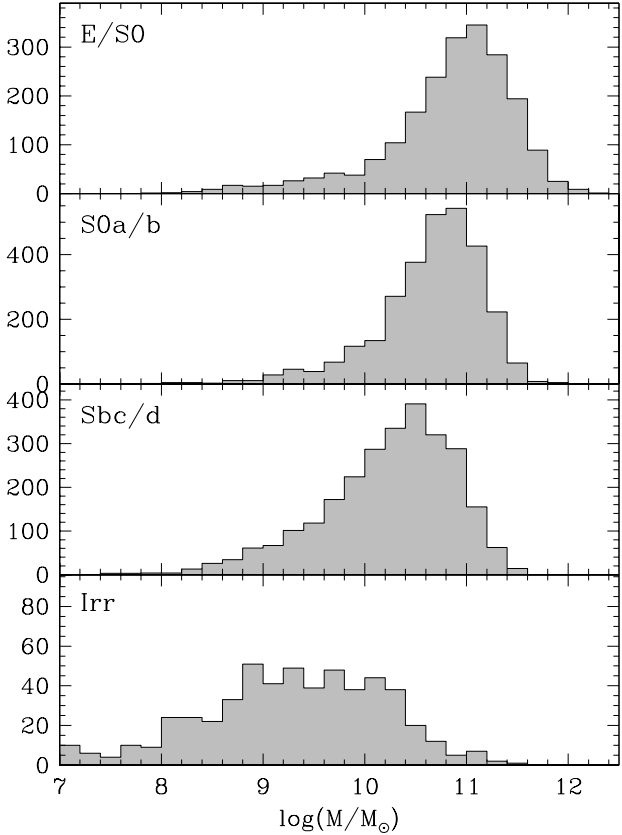}
  \caption{ Distribution of galaxy stellar mass for the different
  morphological classes}
  \label{fig:mass}
  \end{figure}

  \begin{figure*}	
  \centering
  \includegraphics[width=11cm]{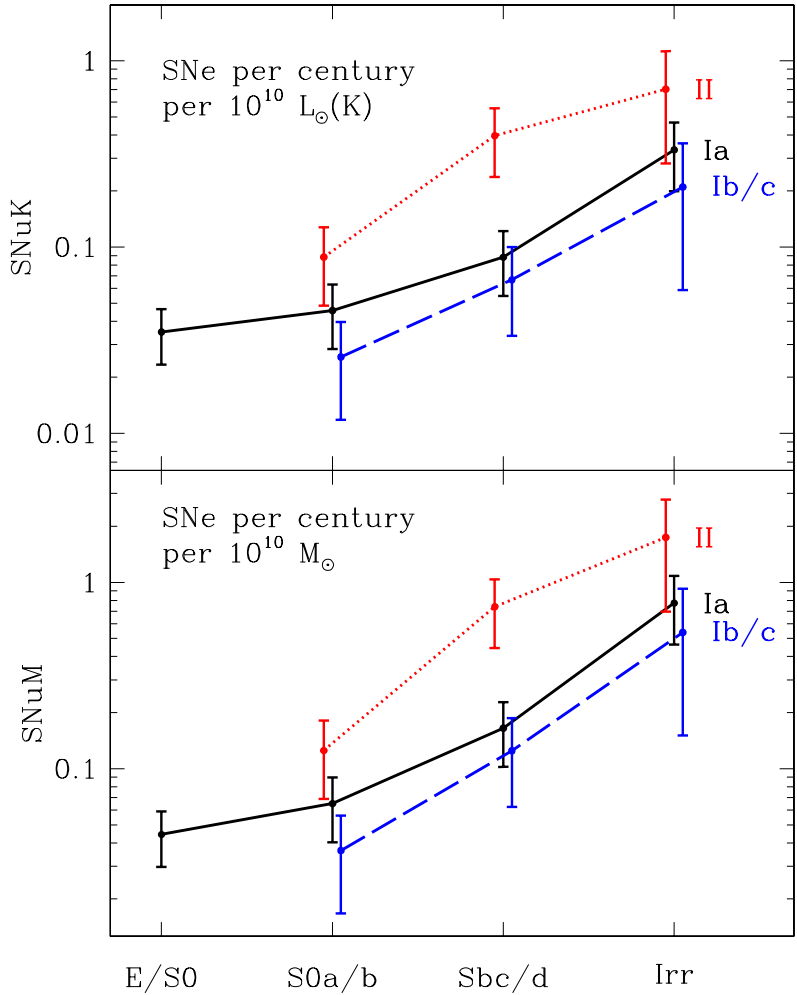}
  \caption{
  {\em Upper panel:} SN rate per K band luminosity as a function of
  morphological index expressed in SNuK (number of SN per century per 
  $10^{10}$ \lsun\ of luminosity in the K band). 
  The lines correspond to type Ia (solid), type II (dotted),
  and Ib/c (dashed) with 1$\sigma$ error bars. 
  {\em Lower panel:} SN rate normalized to the stellar mass and expressed
  in SNuM,
  i.e., number of SNe per century per $10^{10}$ \msun.
  }
  \label{fig:snrate}
  \end{figure*}

The total stellar mass of the galaxies can be derived from the observed
broad-band fluxes by fitting them with spectrophotometric galaxy evolution
models (e.g.,
Brinchmann \& Ellis, 2000; Dickinson et al., 2003). When the redshift
and the morphological type of a galaxy is known, these methods
are expected to give rather accurate results: for a given Initial Mass
Function (IMF), the typical uncertainties in the total masses are less than
40\%.  Larger uncertainties, up to a factor of two, are associated to
the IMF as the low mass stars give a dominant contribution to the
total mass and a secondary one to the luminosity. For example a
difference of a factor of two in galaxy mass is found between the Salpeter
(1955) IMF, which is rich of low-mass stars, and the Kroupa (2001) IMF,
that has a shallower slope for masses below 0.5\msun. Fortunately, as far
as no systematic difference of IMF are present along the Hubble
sequence, this uncertainty has the only effect of shifting up and down
the total mass without introducing relative changes along the Hubble
sequence.

We estimate the mass of the galaxies by using the method developed
by BJ01.
These authors computed the mass-to-light ratio (M/L) by
using galaxy evolutionary synthesis models and found a
tight correlation between M/L and the optical-to-near-IR colors:
for a given K-band magnitude, bluer galaxies are less 
massive because they have a younger populations.
Therefore, the stellar mass can be derived from the K-band luminosity
and the B--K color, which is an indicator of the mean age of the population.
Based on the use of the luminosities in two bands only, this method can be
applied to large samples of galaxies with data available for a limited 
number of filters. This is our case
as most of the galaxies monitored during the SN searches have B and K-band
magnitudes available.

As mentioned above,
the stellar mass derived depends on the adopted IMF which cannot be easily
constrained from the data and therefore must be assumed {\em a priori}.
BJ01 use a Salpeter IMF and scale down of
a factor of 0.7 the resulting M/L
ratio to obtain "maximum disk" mass, i.e., 
the maximum mass compatible to the observed rotation curve.

BJ01 constructed their models to reproduce the
properties of the spiral galaxies. These galaxies show a wide range of
properties, from the quiescent S0 to the very active Sd, and this is
why these models can produce a wide range of colors.
For example, the B-V color
range covered by the BJ01 models goes from 0.29 to 0.95, corresponding,
respectively, to the
average color of the irregulars (B--V=0.27) and ellipticals (B--V=0.96,
Fukugita et al., 1995). As a consequence these models
can be used to compute the M/L ratio along the whole Hubble sequence, 
from Elliptical to Irregulars.

BJ01 tested the robustness of their method by
using several independent evolutionary models, and
studied the effect of introducing secondary bursts
of star formation (galaxy become brighter but bluer) or
dust extinction (galaxies become fainter but redder).
Apart from the effect of the IMF, 
the uncertainties on the M/L of this method
are about 0.4 dex peak-to-peak.

From the coefficients provided in Table~1 of BJ01 we have 
derived the relation
giving the stellar mass from the K-band luminosity $L_K$ and the total color
(B--K):

\begin{equation}
log\left(\frac{M/L_K}{~M_\odot/L_\odot}\right) = 0.212\rm{(B-K)}-0.959
\end{equation}

This equation was used to compute the stellar mass of each galaxy.
The resulting distributions are shown in Figure~\ref{fig:mass} where 
the galaxies are 
divided into four morphological classes.

As a consistency check, we compared the mass distribution 
of our galaxy sample with that 
derived by Kauffmann et al. (2003) for
120.000 local galaxies observed by the Sloan Digital Sky Survey (SDSS). 
These authors use a completely different method based on two stellar absorption
indexes, the 4000\AA\ break and the Balmer line H$\delta$. 
Despite the differences in the selection of the galaxy samples and in the 
methods to estimate the masses, 
the two distributions are remarkably similar: in the SDSS sample, half of the
total stellar mass is contained in galaxies less than
3.1$\times10^{10}$\msun, 
which compares favorably with 3.7$\times10^{10}$\msun\ for 
the present sample.
In both cases the peak of the contribution to the total stellar mass comes
from galaxies of about 6$\times10^{10}$\msun.
This also implies that the present sample is a fair representation of the local
universe as it can be observed in the optical
by modern wide-area surveys as the SDSS.
The results are also consistent with those 
derived by models 
based on a larger number of photometric bands
(e.g., Saracco et al., 2004, Drory et al., 2004).

\section{The SN rate as a function of the morphology of the host galaxy}
\label{sec:snrate}

The SN rates computed both in units of K-band
luminosity and stellar mass are shown in Table~\ref{tab:snrate} and in
figure~\ref{fig:snrate}. 
In the first case the rates are expressed in number of SNe per century
per $10^{10}$ K-band solar luminosity (SNuK), using a
K band magnitude of the Sun of 3.41 (Allen, 1973)
corresponding to a solar luminosity in the K band
of L$_{K,\odot}$=5.67$\times10^{31}$ erg/sec.
In the second case they are expressed in
number of SNe per century and per $10^{10}$ solar 
masses in stars (SNuM).
The errors include both the Poisson errors
due to the SN statistics and the uncertainties of the input parameters
and the bias corrections as explained in C97 and C99.  For the first
time these rates are derived in a fully consistent way by using the
M/L ratios derived from the color of each galaxy and not average 
M/L ratios (Tammann, 1974)
or average colors (Panagia, 2000; Mannucci et al., 2003).

The rates for the Irr galaxies have large uncertainties 
because of the small sample of SNe.
Nevertheless, a clear increase of the
SN rates from E/S0 to S0a/b to Sbc/d to Irr can be seen for all the SN
types. In particular, the lower panel of Fig. 4 shows that all types
of SNe in spirals and Irr
have a remarkably similar behavior, with the rates in the
Irr galaxies being 12-15 times larger than those in the S0a/b galaxies (see
Table~\ref{tab:snrate}).

Qualitatively the result is the same as derived by Mannucci et al. (2003)
by using average colors, but the difference of rates along the Hubble sequence
is even more pronounced:
there the SN Ia rate in the Sbc/d galaxies was found to be
a factor of 2 higher than that in the E/S0, to be compared with the factor of
4 of the present work, and that in the Irr galaxies a factor of 7 higher than
in E/S0, to be compared with the factor of $\sim$17 found here.


  \begin{table}
   \caption{SN rates normalized to the K-band luminosity and to the stellar
   mass. The errors are 1$\sigma$ values and contain the contribution 
   of the
   Poisson statistics of the SN number, often dominant, and of several
   other uncertainties, as explained in C97. The upper limits correspond 
   to 90\% confidence levels.}
  \begin{tabular}{lccc}
  \hline
  \hline
  Type  &  Ia           &   Ib/c          & II \\
  \hline
  \multicolumn{4}{c}{SN rate per K-band luminosity (SNuK)}\\
  \hline
  \vspace*{-2.5mm} \\
  E/S0  &  0.035$_{-0.011}^{+0.013}$ & $<$0.0073          &  $<$0.10          \\
  \vspace*{-2.5mm} \\
  S0a/b &  0.046$_{-0.017}^{+0.019}$ & 0.026$_{-0.013}^{+0.019}$ & 0.088$_{-0.039}^{+0.043}$ \\
  \vspace*{-2.5mm} \\
  Sbc/d &  0.088$_{-0.032}^{+0.035}$ & 0.067$_{-0.032}^{+0.041}$ & 0.40$_{-0.16}^{+0.17}$   \\
  \vspace*{-2.5mm} \\
  Irr   &   0.33$_{-0.13}^{+0.18}$ & 0.21$_{-0.14}^{+0.26}$ & 0.70$_{-0.43}^{+0.57}$   \\
  \vspace*{-2.5mm} \\
  \hline
  \multicolumn{4}{c}{SN rate per Mass (SNuM)}\\
  \hline
  \vspace*{-2.5mm} \\
  E/S0  &  0.044$_{-0.014}^{+0.016}$ & $<$0.0093          & $<$0.013       \\
  \vspace*{-2.5mm} \\
  S0a/b &  0.065$_{-0.025}^{+0.027}$& 0.036$_{-0.018}^{+0.026}$& 0.12$_{-0.054}^{+0.059}$ \\
  \vspace*{-2.5mm} \\
  Sbc/d &  0.17$_{-0.063}^{+0.068}$ & 0.12$_{-0.059}^{+0.074}$ & 0.74$_{-0.30}^{+0.31}$ \\
  \vspace*{-2.5mm} \\
  Irr   &  0.77$_{-0.31}^{+0.42}$ & 0.54$_{-0.38}^{+0.66}$ & 1.7$_{-1.0}^{+1.4}$ \\
  \vspace*{-2.5mm} \\
  \hline
  \end{tabular}
  \label{tab:snrate}
  \end{table}
  
Despite the small number of SNe, the statistical significance of the
difference between the type Ia rate in E/S0 and late spirals/irregulars
is high:
the rate observed in E/S0 would correspond to 0.4 SNe for the 
Irr galaxies, instead of the observed number of 6.8: such a value 
is excluded by the Poisson statistics 
at a significance level higher that 0.9995 
When including also the other sources of errors, 
the final significance remains above 99\%.

\section{The SN rate as a function of the galaxy colors}
\label{sec:sncolor}

  \begin{figure*}	
  \centering
  \includegraphics[width=11cm]{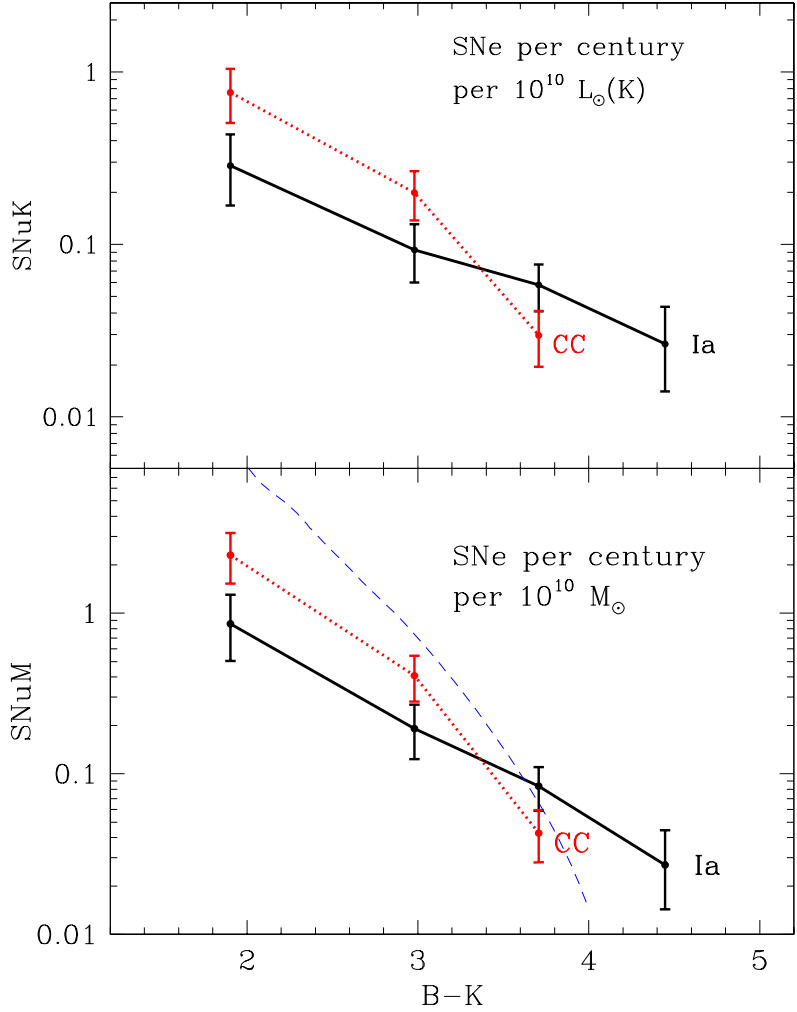}
  \caption{
  {\em Upper panel:} SN rate per K band luminosity 
  expressed in SNuK (number of SN per century per 
  $10^{10}$ \lsun\ of K-band luminosity)
  as a function of the B--K color
  of the parent galaxies.
  The thick lines are
  the results for type Ia (solid) and core-collapse (dotted).
  {\em Lower panel:} SN rate normalized to the stellar mass and expressed
  in SNuM,
  i.e., number of SNe per century per $10^{10}$ \msun\ of stellar mass. 
  The thin dashed line shows
  the relation between the B--K color and the CC SN rate as predicted by a
  model of spiral galaxy (see text) for ages between 
  0.7 and 10 Gyrs after the
  onset of an exponentially declining star formation with e-folding time 
  of 2 Gyr. These two ages correspond, respectively, to Irr and
  E galaxies, whose average colors are B$-$K=2.23 for Irr and B$-$K=3.98 for E
  (Fioc \& Rocca-Volmerange, 1999).
  }
  \label{fig:sncolor}
  \end{figure*}

The SFR is related both to the morphology and to the color of 
the galaxies (e.g., Kennicutt, 1998).
These are not one-to-one relations as the dust content,
the star formation history and the presence of recent mergers
introduce large spreads.
Nevertheless the SFR-color relation is probably tighter than the
SFR-morphology one as the color is more directly 
related to stellar population than morphology. 
For this reason we computed the SN rates also after binning
the galaxies according to their B--K colors. To increase the number of objects
in the each color bin we add up the two types of CC SNe.

The results are shown in Table~\ref{tab:sncolor} and
plotted in Figure~\ref{fig:sncolor}. It is evident
that the rate of the CC SNe has a strong dependence on the galaxy colors, as
already noted, e.g., by C99 when normalizing to the B band. 
The rate of Type Ia SNe is also rapidly
changing with colors, although not as quickly as the for CC SNe. 
This dependence is produced by the normalization to the stellar 
mass as it is not present in C99. 
It also differs from the results by Turatto et al. (1994) which, 
on the base of a limited sample of 5 SNe, studied the 
type Ia SN rate in
early type galaxies, finding no dependence on some
properties of the parent galaxies, such as ISM content in gas or dust.
The variation of the SN rates with the galaxy color
is discussed in detail in the next section.

The thin dashed line in the lower panel of Figure~\ref{fig:sncolor} shows
the relation between the 
B--K color and the CC
SN rate expected from a simple model of spiral galaxy. 
To obtain this, first we computed the expected relation between the B--K color 
and the SFR by using
the GALAXEV models (Bruzual \& Charlot, 2003) with solar metallicity
and an exponentially declining SFR with an e-folding time of 2 Gyrs.
This model can reproduce the properties of a wide range of galaxies,
from early-type spirals to irregular, just by selecting the appropriate
age since the onset of the star formation
(see, for example, Pozzetti et al., 1996).
Second, the SFR was converted into an expected SN rate by integrating
the IMF for masses between 8 and 40 \msun. 
(see, for example, Madau et al., 1998).
As a consequence we don't have any free parameter in plotting this line. 
The model considered is very simple, as
it includes only one star formation history, 
only solar metallicity and no dust. Nevertheless
it is evident from Figure~\ref{fig:sncolor} that
it accurately predicts the behavior of the CC SN rate 
as a function of the galaxy color,
confirming that the progenitors of the CC
SNe are young populations related to the ongoing star formation.  
It is also evident that the number of CC SNe actually observed
is lower by about a factor of 2 than the model predictions.
This is not unexpected as the model considered here
is very simple: as an example,
reducing the metallicity of the galaxies to 40\% solar would
produce galaxies with colors bluer of about B-K$\sim$0.3, greatly
reducing the
discrepancy. Nevertheless, this is also consistent with the
claims from near-IR SN searches (as Mannucci et al., 2003)
that a considerable fraction of the SN expected CC SNe are missing 
from the optical
searches because of the presence of large dust extinctions.

\begin{table}
\caption{Number of SNe, number of galaxies and SN rate
per unit mass in SNuM per color bin. The upper limit is at
90\% confidence level.}
\begin{tabular}{crrrcc}
\hline
\hline
   B--K & Ngal&  Ia &  CC &    SNr(Ia)      & SNr(CC)       \\
\hline
  \vspace*{-2.5mm} \\
$<$2.6  & 1499&  9.0& 20.0&0.86$_{-0.35}^{+0.45}$  & 2.3$_{-0.77}^{+0.86}$\\
  \vspace*{-2.5mm} \\
2.6--3.3& 2178& 15.4& 29.6&0.19$_{-0.07}^{+0.08}$  & 0.41$_{-0.12}^{+0.13}$\\ 
  \vspace*{-2.5mm} \\
3.3--4.1& 3396& 37.3& 17.7&0.084$_{-0.025}^{+0.026}$&0.043$_{-0.015}^{+0.017}$\\
  \vspace*{-2.5mm} \\
$>$4.1  & 1276&  6.0&   0 &0.027$_{-0.013}^{+0.017}$&$<$0.012       \\
  \vspace*{-2.5mm} \\
\hline
\end{tabular}
\label{tab:sncolor}
\end{table}

\section{Discussion}
\label{sec:ia}

  \begin{figure*}
  \centering
  \includegraphics[width=11cm]{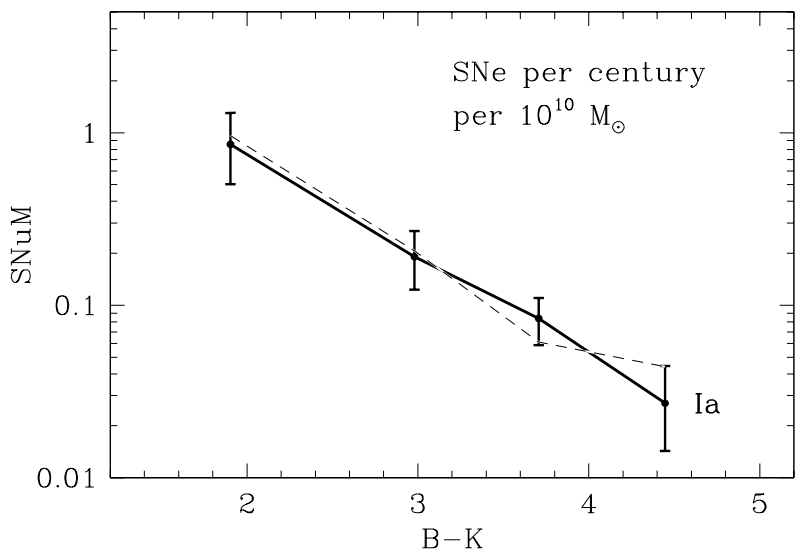}
  \caption{
  The rate of the type Ia SNe (thick solid line) is compared with the 
  result of a toy model (thin dashed line)
  in which the rate is reproduced by a constant value, independent of color
  and fixed at the value measured in the Ellipticals, 
  plus 40\% of the rate of the CC SNe (see Figure~\ref{fig:sncolor}).
  The model well reproduces the observed rate.
  }
  \label{fig:plotIa}
  \end{figure*}

We have used the near-infrared photometry by 2MASS to compute the SN
rates normalized to both the K luminosity and the mass in stars of
the galaxies. We group the parent galaxies either according
to their Hubble types
(Figure~\ref{fig:snrate} and Table~\ref{tab:snrate})
or to their B--K color 
(Figure~\ref{fig:sncolor} and Table~\ref{tab:sncolor}).
A close inspection to these figures reveals
the following facts:

1. The rate of type Ia, Ib/c and II SNe per unit of mass increases by
   factors of 2, 3 and 5 respectively from early to late
   spirals. 
   This fact is universally accepted for the CC SNe
   whose rates are known 
   to be closely related to the SFR and not to the total stellar mass. 
   For type~Ia SNe this is a more controversial point, 
   as discussed below.

2. The behavior of the mass-normalized rates of various SN classes are
   remarkably similar, as shown by the slopes of the curves in the
   bottom panels of Figure~\ref{fig:snrate} and \ref{fig:sncolor}.
   Apparently this finding is at odds with the standard picture for
   the progenitors of type Ia SNe. The parent binary system could be either
   single-degenerate or double-degenerate (e.g., Woosley \& Weaver,
   1986; Wheeler \& Harkness, 1993; Branch et al., 1995; 2001;
   Hoflich et al. 1995;, Yungelson \& Livio 2000, Nomoto 2003) and
   both cases suggest an evolved progenitor. Our result indicates that
   an ``evolved'' binary system is not synonymous of ``very old''
   binary system. On the contrary Figures~\ref{fig:snrate} and
   \ref{fig:sncolor} suggest that most progenitors of type Ia SNe
   occurring in late Spirals/Irr are associated with a relatively
   young stellar component (cf. Greggio \& Renzini 1983).  
   We note that this scenario was
   already proposed on theoretical grounds by Dallaporta in the early
   70's (Dallaporta 1973).
   
   Our results can also be expressed in terms of the delay time
   between a newly born white dwarf (i.e. the primary) and the 
   the SN explosion (indicated by $\tau$ in the parametrization
   by Madau et al., 1998).
   Since the time
   scale of the stellar evolution for stars near 8\msun\ is about
   50 Myr, our results lend support to the models of progenitors 
   characterized by `short' delay time. This point will be discussed 
   in detail
   in a forthcoming paper (Mannucci et al., 2004, in preparation).

3. The rate of type Ia SNe per unit mass increases by a factor of
   about 4 from E/S0 to Sbc/d and up to a factor of about 17 in Irr
   galaxies. 
   The same effect can also be seen when binning the galaxies according to
   their colors: the ratio between the type Ia SN rate in galaxies bluer than
   B--K=2.6 and redder than B--K=4.1 is larger than 30. 
   Both these results have a very high statistical significance (larger
   than 0.99 confidence level). 

   The existence of an order of magnitude difference in the rates
   between late Spirals/Irregulars and Ellipticals implies that the
   frequency of binary systems exploding as a type Ia SN per unit time
   changes appreciably when the parent population ages. This is not totally
   unexpected in the theoretical framework of the double degenerate
   scenario. For example Livio (2001) (see also Yungelson et al. 1994)
   predicts a decrease of a factor of 10 of the merger rate in ``old''
   ($10^{10}$ yr) stellar population with respect to the ``young''
   ($10^8$ yr) stellar component.

4. However, this might represent only one side of the coin. The
   aging of the parent population affects not only the SN rate but
   also the properties of the SN explosions.  Li et al. (2001) have
   shown that the class of the Ia SNe are characterized by a large number of
   ``peculiar'' objects (36\%$\pm 9$\%). This high degree of
   inhomogeneity reflects the existence of systematic differences in
   the properties of the SN explosion, such as the velocity of the
   ejecta (Filippenko, 1989; Branch \& van den Bergh, 1993, Nugent et
   al.  1995), the luminosity and photometric evolution (Della Valle
   \& Panagia, 1992; Phillips, 1993; Hamuy et al., 1996, 2000; Canal,
   Ruiz-Lapuente \& Burkert, 1996; Howell, 2001), and the rate of
   occurrence (Della Valle \& Livio, 1994).  Howell (2001) has shown
   that underluminous SNe-Ia, like 1991bg (Filippenko et al., 1992;
   Turatto et al., 1996) or SN 1992K (Hamuy et al., 1994) are twice
   more common in early type galaxies than in late-type galaxies,
   whereas overluminous SNe-Ia are more common in late-type galaxies
   than in early type galaxies (15 vs. 2).  Our results are
   consistent with a picture in which the underluminous, 1991~bg-like
   type Ia SNe are preferentially associated to an old stellar
   population, whereas overluminous type Ia SNe tend to occur in
   spiral and star-forming galaxies.

5. The SN rate as a function of the parent galaxy color in
   Figure~\ref{fig:sncolor} can be used to estimate the relative
   contribution of the ``old'' and ``young'' channels.  The
   behavior of the SN rate can 
   be reproduced by combining a constant contribution (i.e.,
   independent of galaxy colors) due to ``old'' progenitors, plus a
   contribution proportional to the SFR due to the ``young''
   progenitors.  The first contribution must be similar to the value
   measured in the reddest (i.e., more quiescent) galaxies or in the
   Ellipticals, while the other contribution must show a behavior
   similar to that of the CC SNe.  The result of this simple model is
   shown in Figure~\ref{fig:plotIa}: the type Ia rate is well
   reproduced by the constant rate of 0.044 SNuM measured in the E/S0
   plus a fixed fraction (40\%) of the rate of the CC SNe.

   The best fitting agreement between the two curves is obtained for:
   \begin{equation}
   {\rm SNr(Ia)} =  (0.047\pm0.009) + (0.35\pm0.08)\cdot{\rm SNr(CC)}
   \label{eq:Ia}
   \end{equation}
   where the errors are the formal fitting
   uncertainties on the two free parameters.
   These values can also reproduce the
   type Ia SN rate as a function of the morphological type in
   Figure~\ref{fig:snrate} with great accuracy.

   The first term is due to the ``old'' progenitors, responsible of
   most SNe-Ia in the Ellipticals, about 50\% in S0a/b, about 20\% in
   Sbc/d and a few \% in the Irr.  The second term is roughly
   proportional to the SFR (assuming that the CC SN rate tracks
   the instantaneous SFR) and dominates in the late-type
   galaxies. 

   We note that the type Ia SN rate in perfectly quiescent galaxies,
   corresponding to the first (constant) term in eq.~\ref{eq:Ia},
   could be even lower.  Indeed, the type Ia SN rate in radio-loud
   early type galaxies appears to be enhanced, by a factor $\sim 4$,
   with respect to the radio-quiet sample (Della Valle \& Panagia, 2003;
   Della Valle et al., 2004). 
   The latter ($0.023^{+0.012}_{-0.008}$) is similar to the
   value measured here for the reddest galaxies $\sim$0.027 SNuM, (see
   Table~\ref{tab:sncolor}). In conclusion, all of this 
   provides (additional) empirical support to the idea, proposed in the
   past years (Della Valle \& Livio 1994; Ruiz-Lapuente, Burkert, \&
   Canal 1995) that SNe Ia in late- and early-type galaxies may be
   originated by different types of progenitors and/or through
   different explosive channels.



\end{document}